\documentclass[reprint,
 amsmath,amssymb,
 aps,
]{revtex4-2}

\usepackage{graphicx}% Include figure files
\usepackage{dcolumn}% Align table columns on decimal point
\usepackage{bm}% bold math
\usepackage{siunitx}
\usepackage{cleveref}
\usepackage{caption}
\usepackage{subcaption}
\usepackage{natbib}
\begin{document}

%\preprint{APS/123-QED}

\title{Integer vs. half-integer spin on an approximate honeycomb lattice}% Force line breaks with \\

\author{V. J. Stewart$^{1,2}$}
\author{J. R. Chamorro$^{1,2}$}
\author{T. M. McQueen$^{1,2,3}$}
\email{mcqueen@jhu.edu}
\affiliation{
 $^1$Department of Chemistry, The Johns Hopkins University, Baltimore, Maryland 21218, USA\\
 $^2$Institute for Quantum Matter, Department of Physics and Astronomy, The Johns Hopkins University, Baltimore, Maryland 21218, USA \\
 $^3$Department of Materials Science and Engineering, The Johns Hopkins University, Baltimore, Maryland 21218, USA
}

\date{\today}% It is always \today, today,

\begin{abstract}
Recent interest in honeycomb lattice materials has focused on their potential to host quantum spin liquid (QSL) states. Variations in bond angles and spin allow a range of interesting behaviors on this lattice, from the predicted QSL ground state of the Kitaev model to exotic magnetic orders. Here we report the physical properties of two compounds with rare earths on an approximate honeycomb lattice. The isostructural compounds Nd$_2$S$_5$Sn (J = $\frac{9}{2}$) and Pr$_2$S$_5$Sn (J\,=\,4) permit a direct comparison of half-integer versus integer spins on this lattice. We find strikingly different magnetic properties for the two compounds. Nd$_2$S$_5$Sn orders antiferromagnetically at T$_N$\,$\approx$\,2.5\,K, and undergoes several magnetic transitions to other ordered states under applied field.  Pr$_2$S$_5$Sn displays no magnetic ordering transition above T\,=\,0.41 K, and may be proximate to a spin liquid state.
\end{abstract}

\maketitle

%\tableofcontents

\section{\label{sec:level1}Introduction}

Honeycomb lattice materials have been of recent interest as quantum spin liquid (QSL) candidates, as they can host magnetically frustrated spin configurations that may have a disordered ground state\,\cite{balents, broholm, jcqsl}. Much of this interest has arisen due to the Kitaev model, which predicts a quantum spin liquid ground state on a honeycomb lattice with the right magnetic exchange interactions, and is exactly solvable for S\,=\,$\frac{1}{2}$\,\cite{kitaev}. Candidate materials for this Kitaev spin liquid include $\alpha$-RuCl$_3$ and Ir$^{4+}$ honeycomb iridates such as Li$_2$IrO$_3$ and Na$_2$IrO$_3$\,\cite{jcqsl, sears}. So far, all of these candidate materials have been found to magnetically order, but the unconventional magnetic orders they adopt suggest that they are adjacent to a QSL state\,\cite{sears, chal, singh}.

Looking at honeycomb materials beyond the prototypical spin $\frac{1}{2}$ on an ideally symmetrical lattice is also valuable. The potential of larger spins to allow a QSL state has been sometimes been investigated. Higher-spin models cannot be solved exactly, and have weaker quantum fluctuations than S\,=\,$\frac{1}{2}$. Even so, computational studies of S\,=\,1 moments with both Kitaev and Heisenberg interactions predict a spin liquid region of the phase diagram if the Heisenberg/Kitaev exchange ratio is appropriate\,\cite{stavro, dong, bishop}. A$_3$Ni$_2$XO$_6$ with X=Bi, Sb and A=Li, Na have been suggested as candidate materials\,\cite{stavro}.

Extensions to the model with a bond-dependent off-diagonal exchange term included along with Kitaev and Heisenberg terms in the Hamiltonian have been proposed to explain the magnetic order seen in Na$_2$IrO$_3$\,\cite{rau}. In this material, the absence of global hexagonal or trigonal symmetry allows the Ir-O-Ir bond angles to deviate from 90$^\circ$. Although this may move the material away from a Kitaev spin liquid state, it allows study of the relationship between this state and the long-range magnetic orders adopted. 

Even in the absence of Kitaev interactions, spins on a honeycomb lattice can display a range of exotic magnetic states\,\cite{rao, venderbos}. Further, recent studies have shown proximal spin liquid behavors in a number of layered rare earth compounds, including NaYbX$_2$ (X=O, Se) and YbMgGaO$_4$, as well as 3D variants including Ce$_2$Zr$_2$O$_7$ and Pr$_2$Zr$_2$O$_7$\,\cite{NaYbO2, YbMgGaO4, Ce2Zr2O7, Pr2Zr2O7}.

Here we present magnetic and thermodynamic characterization of Nd$_2$S$_5$Sn and Pr$_2$S$_5$Sn, two isostructural materials containing an approximate honeycomb lattice of rare earth ions. They allow a direct comparison between integer (\textit{f2}\,Pr$^{3+}$) and half-integer (\textit{f3}\,Nd$^{3+}$) spins on this lattice. Strikingly, despite point charge calculations revealing a very similar single ion ground state, different physical properties result. The half-integer spin Nd$_2$S$_5$Sn orders antiferromagnetically at 2.5\,K, and undergoes a series of transitions under applied field, adopting an intermediate magnetic order between its antiferromagnetic and ferromagnetic states. In integer-spin  Pr$_2$S$_5$Sn on the other hand, no magnetic ordering is observed down to 0.41\,K. These results add to our understanding of the complex magnetic behavior seen in honeycomb materials.

\begin{figure}

\captionsetup[subfigure]{labelformat=empty}
\begin{subfigure}{8.6cm}

\includegraphics[width=\textwidth, keepaspectratio]{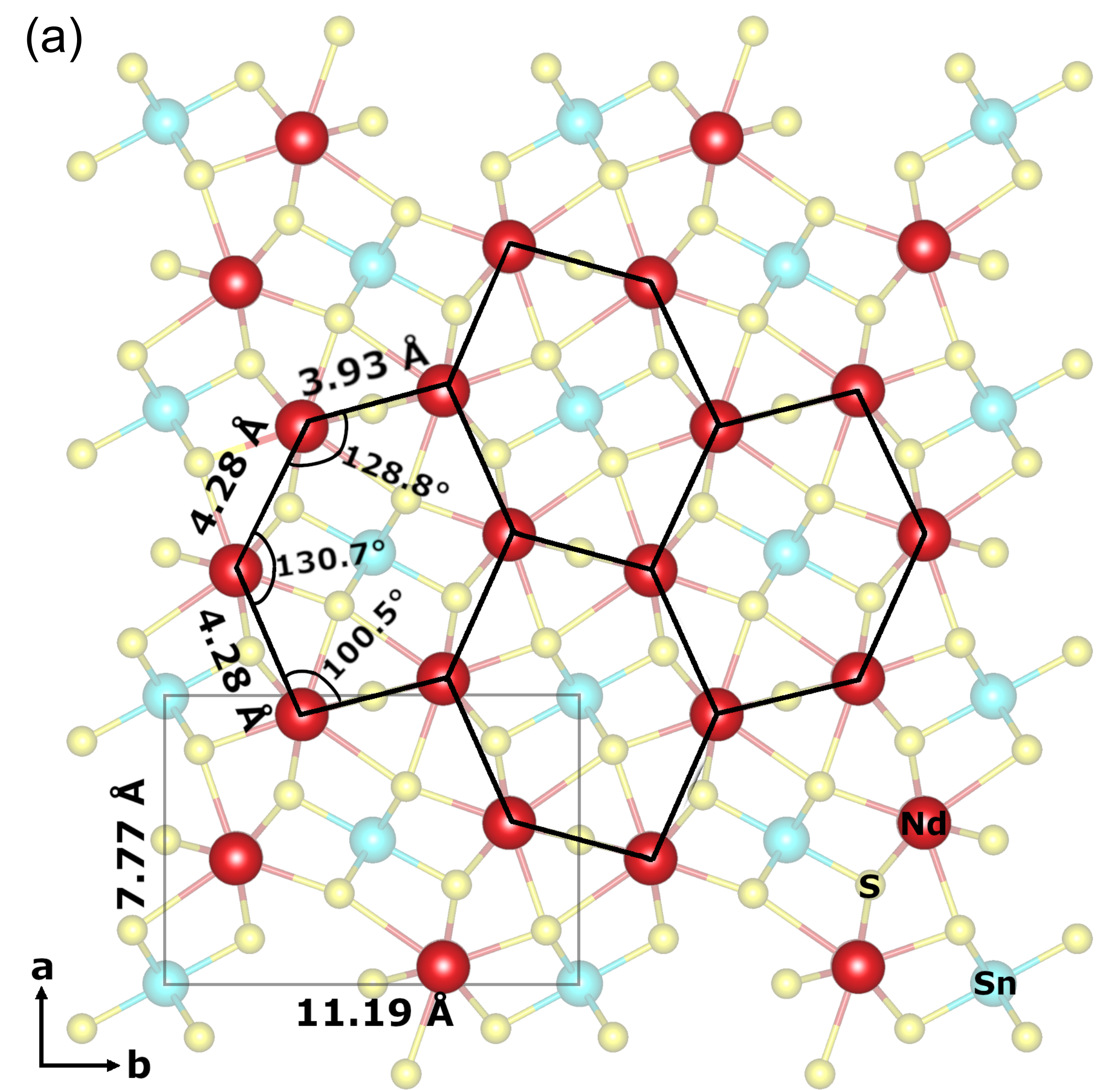}
\caption{\label{fig:fig1a}}
\end{subfigure}
\hfill
\begin{subfigure}{8.6cm}
\includegraphics[width=\textwidth, keepaspectratio]{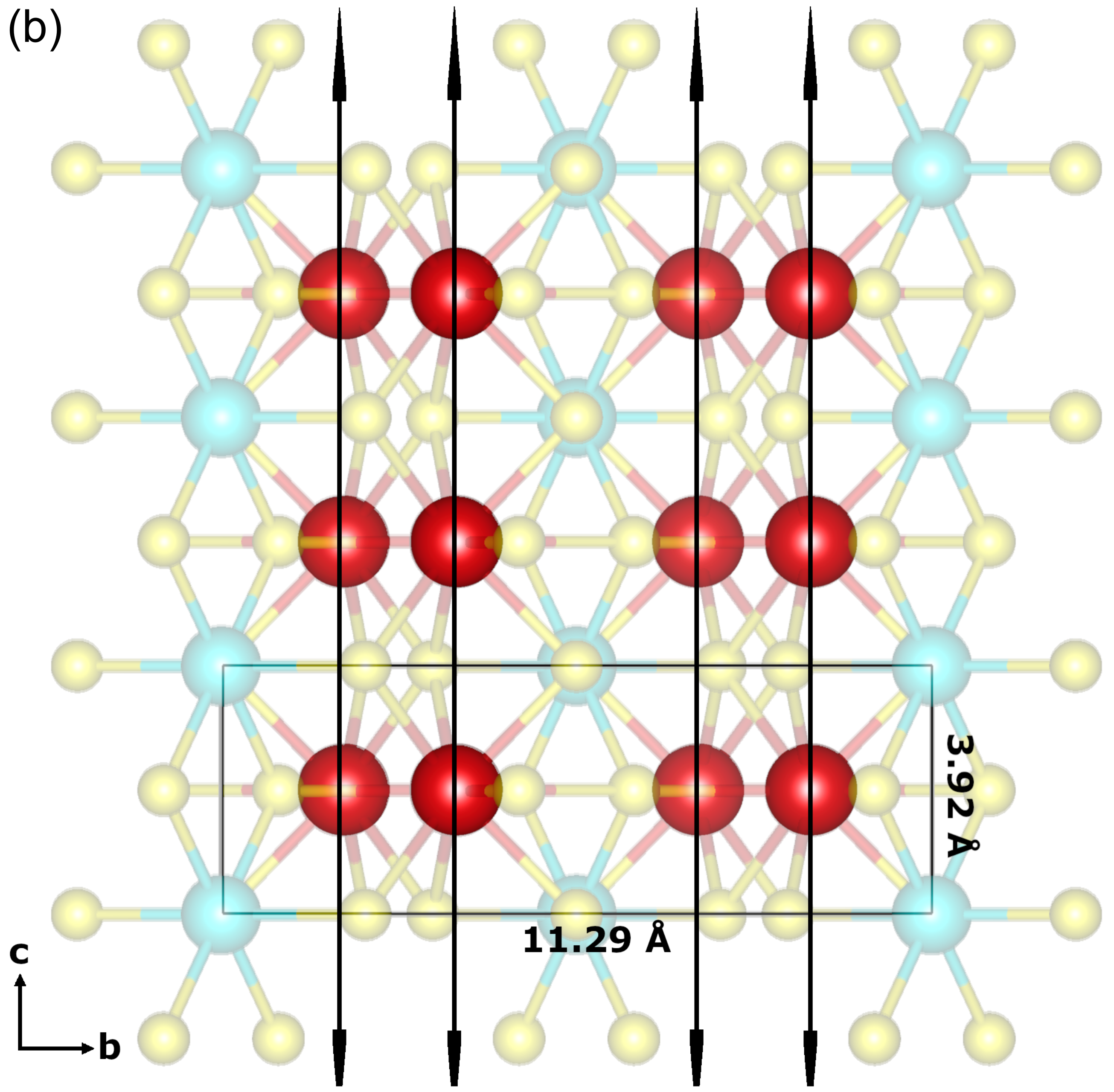}
\caption{\label{fig:fig1b}}
\end{subfigure}{}
\caption{\label{fig:fig1}(a) The structure of Nd$_2$S$_5$Sn in the \textit{ab} plane, showing the approximate honeycomb lattice of Nd${^{3+}}$ ions. Lattice parameters and bond lengths were estimated by refinement of powder x-ray diffraction data in space group Pbam. Nd atoms are shown by red spheres, Sn by cyan, and S by yellow. For the isostructual Pr$_2$S$_5$Sn, a\,=\,7.77\,\AA, b\,=\,11.23\,\AA, and c\,=\,3.95\,\AA. The Pr-Pr distances along the hexagons are 3.94\,\AA, 4.29\,\AA, 4.29\,\AA , and their internal angles are 129.8$^\circ$, 130.0$^\circ$, 100.1$^\circ$. (b) The structure in the \textit{bc} plane, showing the 1D columns of Nd${^{3+}}$.}
\end{figure}

\section{\label{sec:level1}Methods}

Nd$_2$S$_5$Sn, Pr$_2$S$_5$Sn, and a non-magnetic analogue La$_2$S$_5$Sn were prepared from stoichiometric ratios of the elements. Starting materials were sealed in quartz tubes under $\approx\,$0.2\,bar argon gas and heated at 870\,K for 4 hr. After cooling and regrinding, pellets of the materials in evacuated quartz tubes were heated at a rate of 100\,K/hr to 1320\,K. After 12 hours, they were cooled to 870\,K at a rate of 15\,K/hr and water-quenched. Air-stable grey powders were obtained. Products were checked with x-ray diffraction, and if necessary additional sulfur was added to the sample and the 1320\,K heating cycle was repeated.

X-ray diffraction patterns were collected on a laboratory Bruker D8 Focus diffractometer (Cu tube, K$\alpha$1\,=\,1.540596\,\AA , K$\alpha$2\,=\,1.544493\,\AA ) with a LynxEye detector. Structural refinements were performed with \textsc{GSAS-II}\,\cite{gsas}. Structures were visualized with \textsc{Vesta}\,\cite{vesta}. The crystal field splitting for a point charge model of Nd$^{3+}$ and Pr$^{3+}$ was computed using \textsc{PyCrystalField}\,\cite{pycrystalfield}.

Magnetization data were collected on a Quantum Design Physical Property Measurement System (PPMS) using the ACMS option,  and on a Quantum Design Magnetic Property Measurement System (MPMS). Magnetic susceptibility was approximated as magnetization divided by the applied magnetic field ($\chi\,\approx\,$ M/H). Heat capacity data were collected on the PPMS using the semi-adiabatic method and a 1\% temperature rise. The low-temperature heat capacity of Nd$_2$S$_5$Sn was additionally measured using a long-pulse method with 30\% temperature rise, and analyzed using the \textsc{LongPulseHC} software package\,\cite{LPHC}.

\section{\label{sec:level1}Results}

\subsection{\label{sec:level2}Structure}

Both compounds were refined in the space group \textit{Pbam}, consistant with the literature. The structural parameters obtained were also consistent with previous reports\,\cite{jaulmes, dasz}. Refinement indicated a small Ln$_{10}$OS$_{14}$ (Ln = Nd, Pr) impurity in each compound (estimated weight fraction: 3.0\% for Nd, 2.9\% for Pr).

The Ln$^{3+}$ rare earth ions of Pr$_2$S$_5$Sn form an approximate honeycomb lattice in the ab plane. Each hexagon of Ln$^{3+}$  is skewed away from equilateral, as shown in Figure~\ref{fig:fig1a}. Along the c direction, the Ln$^{3+}$ ions align to form a column (Figure~\ref{fig:fig1b}).  Each rare earth is coordinated by nine sulfur atoms, while tin and sulfur are respectively octahedrally and tetrahedrally coordinated. This structure may allow Ln$_2$S$_5$Sn to behave as a pseudo-two-dimensional crystal, with each column (a 1D\,chain) functioning as a single unit for magnetic exchange. In both componds, the distances between nearest-neighbor Ln$^{3+}$ within the planes and along the columns are similar, ranging from 3.9 to 4.3\,\AA. All Ln$^{3+}$ atoms are connected via sulfur bonds, and the presence of this bonding between layers makes the stacking fault disorder present in some layered honeycomb materials unlikely here.

\begin{figure}
\includegraphics[width=8.6cm, keepaspectratio]{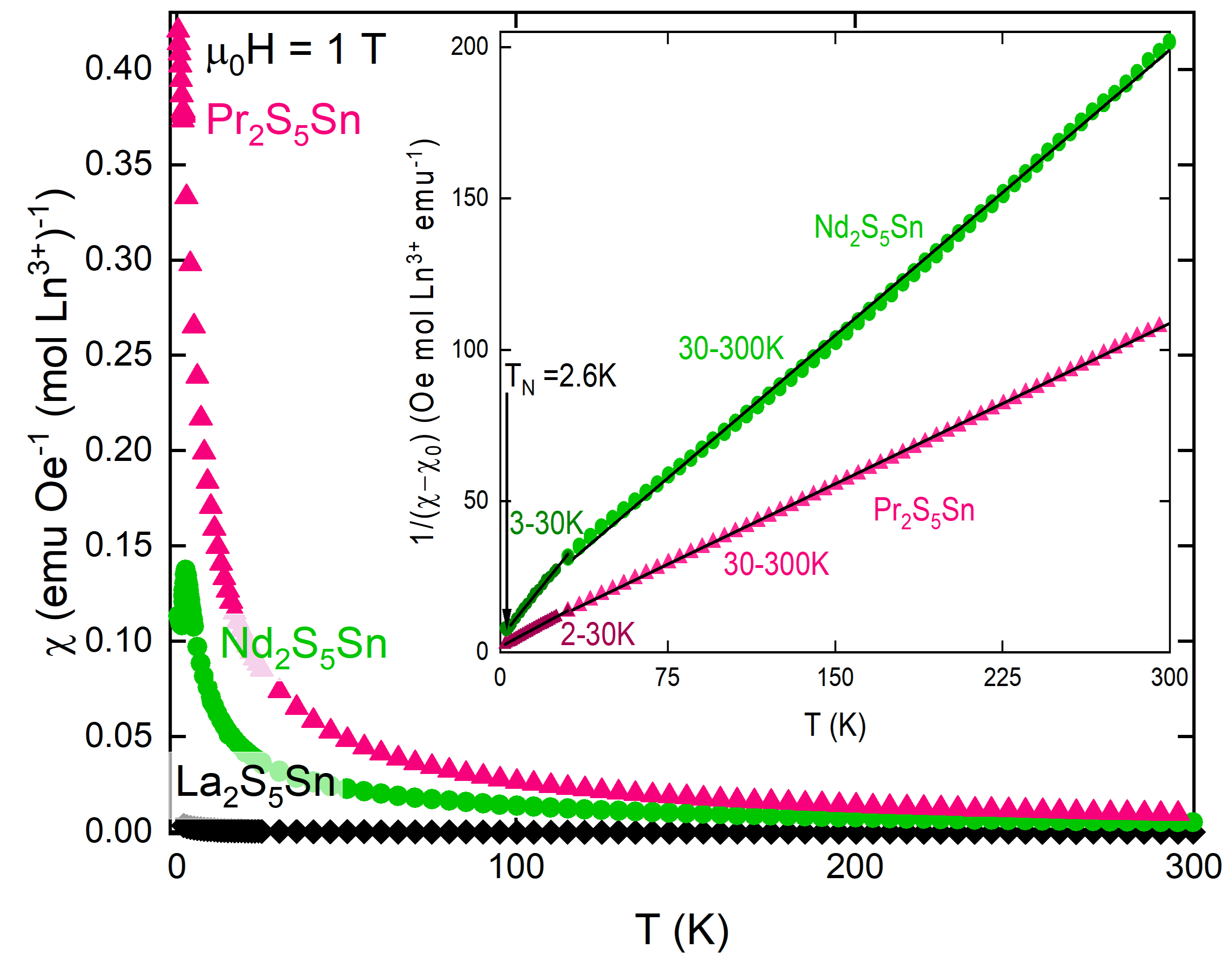}
\caption{\label{fig:fig2}Magnetization versus temperature for Nd$_2$S$_5$Sn (green circles) and Pr$_2$S$_5$Sn (pink triangles). The inset shows Curie-Weiss fits to high and low temperature regions for each compound. The non-magnetic analog La$_2$S$_5$Sn (black diamonds) is also included for reference.}
\end{figure}

\subsection{\label{sec:level2}Magnetization}

\begin{ruledtabular}
\label{tab:table1}
\begin{table}[b]
\caption{\label{tab:table1}Parameters obtained from Curie-Weiss analysis of Nd$_2$S$_5$Sn and Pr$_2$S$_5$Sn magnetization data. Low temperature (LT) and high temperature (HT) ranges were fitted separately. The units of the Curie constant c are \si{emu.K.(Oe.mol.Ln^{3+}})$^{-1}$}.
\begin{tabular}{ccccc}
&
Pr LT &
Pr HT &
Nd LT &
Nd HT \\
\colrule
Range (K) & 2-30 & 30-300 & 3-30 & 30-300 \\
c   & 2.529(4) & 2.825(1) & 1.093(7) & 1.589(3) \\
$\theta$ (K) & -4.7(1) & -7.3(8)& -5.3(5) & -16(5) \\
p$_{eff}$ ($\mu_B$) & 4.49(2) & 4.75(1) & 2.96(2) & 3.57(1) \\
\end{tabular}
\end{table}
\end{ruledtabular}

Magnetization versus temperature measurements (M(T)) show a clear anti-ferromagnetic phase transition for Nd$_2$S$_5$Sn near T$_{N}$\,=\,2.6\,K, while Pr$_2$S$_5$Sn appears paramagnetic down to T\,=\,0.41\,K (Figure~\ref{fig:fig2}). Parameters obtained from Curie-Weiss fits for each compound are given in Table~\ref{tab:table1}. Fits were performed over the range T\,=\,30\,-\,300\,K, as well as over a lower temperature range (3\,-\,30\,K for Nd$_2$S$_5$Sn and 2\,-\,30\,K for Pr$_2$S$_5$Sn) to avoid excited crystal fields. In all cases, best fit was achieved with the Pauli paramagnetic susceptibility set at $\chi_0$\,=\,0. The room temperature susceptibility of the non-magnetic analog La$_2$S$_5$Sn was $\chi$\,=\,-2.06$\cdot$10$^{-4}$ \si{emu.(Oe.mol.Ln^{3+}})$^{-1}$. By comparison to the literature diamagnetic susceptibility of La$^{3+}$ ($\chi_D$\,=\,-2$\cdot$10$^{-5}$ \si{emu.(Oe.mol.Ln^{3+}})$^{-1}$), this value is consistant with a negligible $\chi_0$\,\cite{bain}. Over the 30\,-\,300\,K range, the Weiss temperature of the Nd was estimated as $\theta_w$\,=\,-16(5)\,K, while for the Pr it was $\theta_w$\,=\,-7.3(8)\,K. These $\theta_w$ values indicate that antiferromagnetic interactions are dominant and that the interaction strength is larger in the Nd compound. The effective magnetic moment calculated from the Curie constant for this temperature range is p$_{eff}$\,=\,3.57(1)\,$\mu_B$ for the Nd compound and p$_{eff}$\,=\,4.75(1)\,$\mu_B$ for the Pr; the Nd is slightly lower than the free-ion magnetic moment (3.62\,$\mu_B$) and the Pr is somewhat higher than the free-ion moment (3.58\,$\mu_B$).

For the lower temperature range, p$_{eff}$\,=\,4.49(2) for Pr and p$_{eff}$\,=\,2.96(2) for Nd. We can compare these to the low-temperature moments for the Pr and Nd pyrochlores, which are also magnetically frustrated and have properties significantly influenced by their crystal field states: Pr$_2$Pb$_2$O$_7$ (p$_{eff}$\,=\,2.53(1)\,$\mu_B$), Pr$_2$Zr$_2$O$_7$ (p$_{eff}$\,=\,2.5(1)\,$\mu_B$), Pr$_2$Sn$_2$O$_7$ (p$_{eff}$\,=\,2.6\,$\mu_B$),  Nd$_2$Pb$_2$O$_7$ (p$_{eff}$\,=\,2.55(7)\,$\mu_B$), Nd$_2$Zr$_2$O$_7$ (p$_{eff}$\,=\,2.543(2)\,$\mu_B$), and Nd$_2$Sn$_2$O$_7$ (p$_{eff}$\,=\,2.63 (3)\,$\mu_B$\,\cite{hallas, Kimura2013, princep, matsuhira, CiomagaHatnean2014, Bertin}.

\begin{figure}
\includegraphics[width=8.6cm, keepaspectratio]{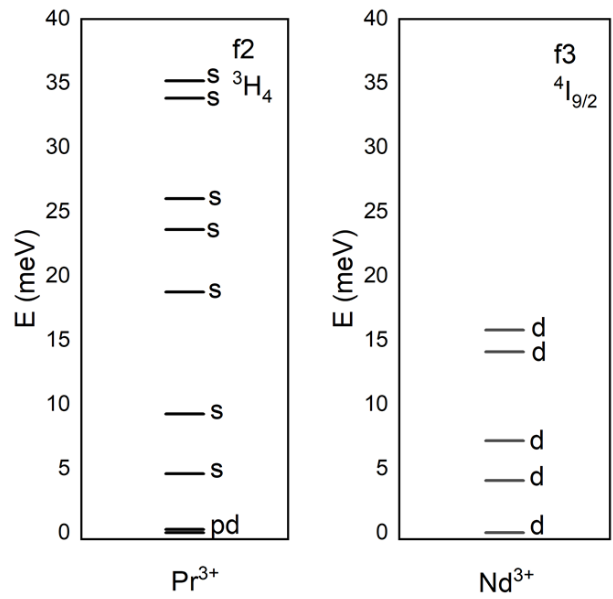}
\caption{\label{fig:fig3}Computed single-ion crystal field levels for Pr$^{3+}$ and Nd$^{3+}$. s indicates a singlet state, d a doublet, and pd a “pseudo-doublet”. The low-energy pseudo-doublet of Pr$^{3+}$ can explain its paramagnetic behavior.}
\end{figure}

Crystal field splittings computed from the point charge model offer an explanation for these results (Figure~\ref{fig:fig3}). Nd$^{3+}$ (J\,=\,$\frac{9}{2}$) splits into five Kramers doublets, while Pr$^{3+}$(J\,=\,4) splits into nine singlet states. In Pr$^{3+}$, the energy gap between the two lowest states is only 0.27\,meV ($\approx\,2.6\,K$). Due to this low energy barrier, these states may act as a “pseudo-doublet”, allowing an effective J\,=\,$\frac{1}{2}$ and providing the unpaired spins necessary for the paramagnetic behavior of Pr$_2$S$_5$Sn. 

\begin{figure*}
\captionsetup[subfigure]{labelformat=empty}
\begin{subfigure}{8.6cm}
\includegraphics[width=\textwidth, keepaspectratio]{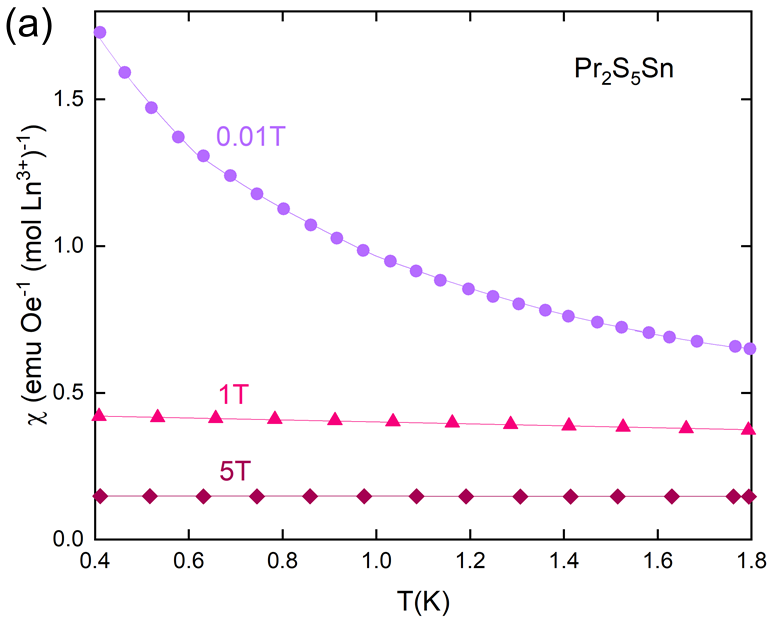}
\caption{\label{fig:fig4a}}
\end{subfigure}
\hfill
\begin{subfigure}{8.6cm}
\includegraphics[width=\textwidth, keepaspectratio]{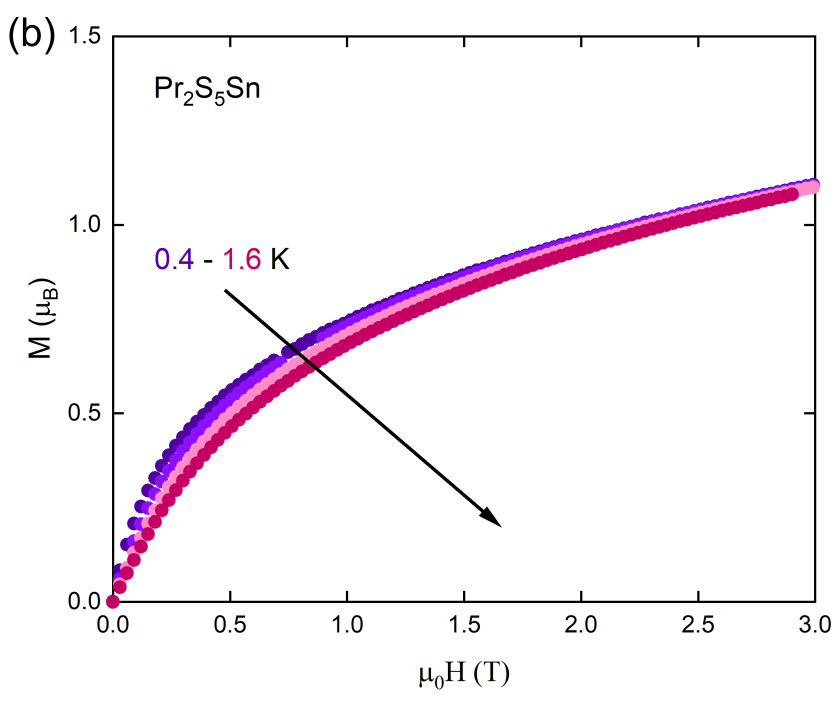}
\caption{\label{fig:fig4b}}
\end{subfigure}{}
\caption{\label{fig:fig4}(a) Magnetization versus temperature for Pr$_2$S$_5$Sn, measured from T\,=\,0.4\,-\,1.8 K in a $^3$He system. No ordering transition was observed. (b) Magnetization versus field for Pr$_2$S$_5$Sn.}
\end{figure*}

\begin{figure*}
\captionsetup[subfigure]{labelformat=empty}
\begin{subfigure}{8.6cm}
\includegraphics[width=\textwidth, keepaspectratio]{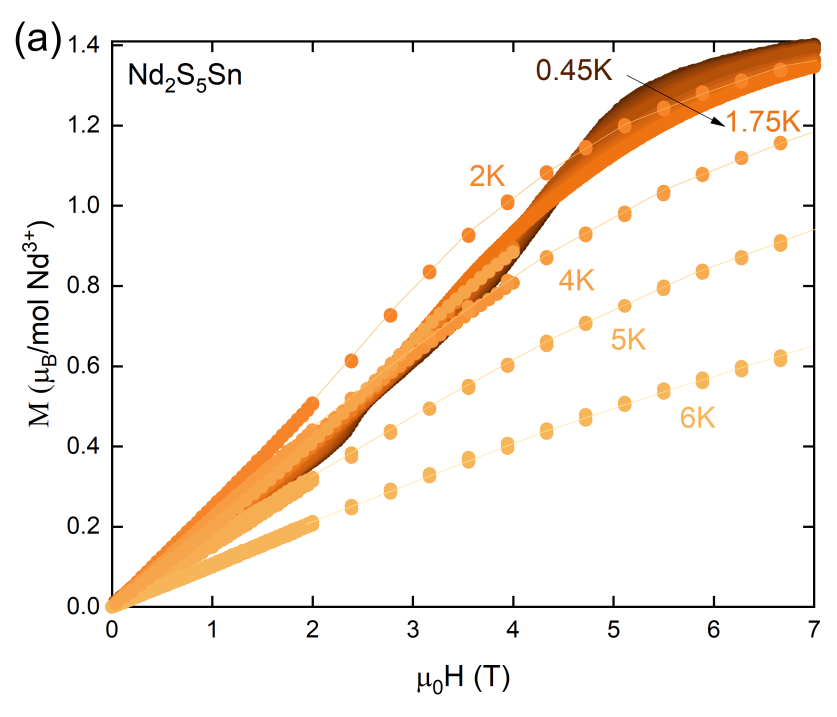}
\caption{\label{fig:fig5a}}
\end{subfigure}
\hfill
\begin{subfigure}{8.6cm}
\includegraphics[width=\textwidth, keepaspectratio]{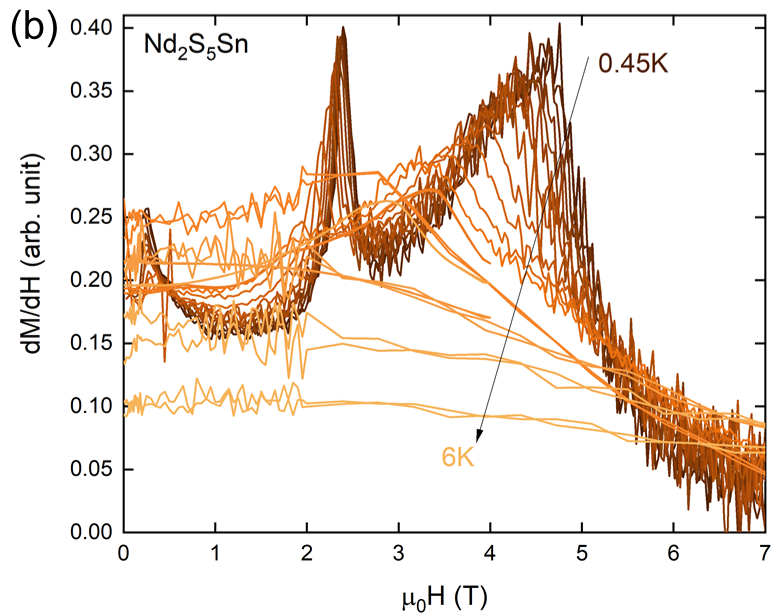}
\caption{\label{fig:fig5b}}
\end{subfigure}{}
\caption{\label{fig:fig5}(a) Magnetization versus field for Nd$_2$S$_5$Sn at temperatures from T\,=\,0.45\,-\,6 K. No hysteresis was observed in field sweeps. (b) Derivative of Nd$_2$S$_5$Sn magnetization vs field, showing three distinct peaks at temperatures $\leq$ 1.8\,K.}
\end{figure*}

To investigate possible ordering in Pr$_2$S$_5$Sn at T\,$<$\,2\,K, M(T) and magnetization versus field (M(H)) measurements were performed in a $^3$He system (Figure~\ref{fig:fig4}). No evidence of magnetic ordering was found down to T\,=\,0.41 K, either in M(T) or in M(H) at any field. We thus conclude that Pr$_2$S$_5$Sn remains paramagnetic for  T\,$>$\,0.41\, K. The M(H) curves approach field saturation as expected for a paramagnet at low temperatures, but do not appear to fully saturate in the 7\,T range measured.

To look more closely at the observed phase transition in Nd$_2$S$_5$Sn, M(H) data were collected at temperatures between 0.45 and 6\,K (Figure~\ref{fig:fig5}). Data points with temperature or sample center position values outside of two standard deviations are excluded from the figure. No hysteresis was observed. Derivatives of the M(H) curves allow clear visualization of the features of this data. At T\,=\,3\,K and above, as expected the M versus H curves are smooth and featureless, consistent with the absence of the phase transition at these temperatures. 

\begin{figure*}

\captionsetup[subfigure]{labelformat=empty}
\begin{subfigure}{8.6cm}
\includegraphics[width=\textwidth, keepaspectratio]{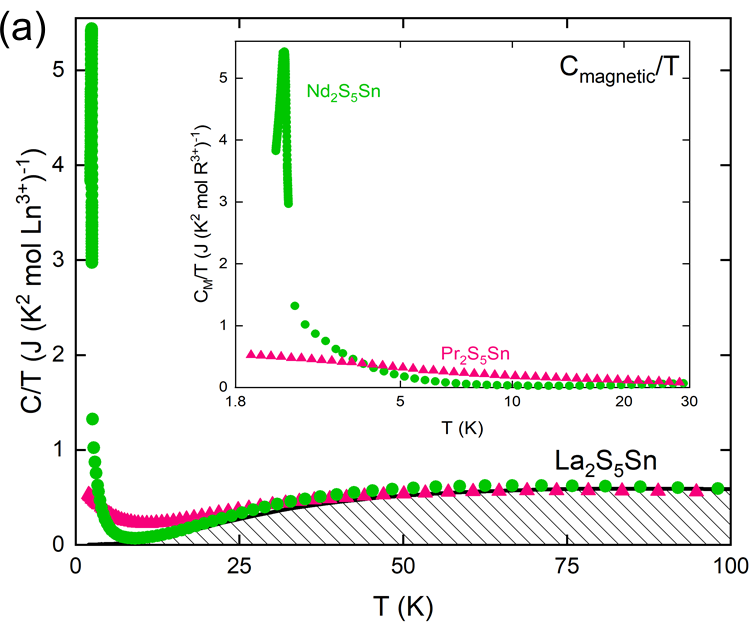}
\caption{\label{fig:fig6a}}
\end{subfigure}
\hfill
\begin{subfigure}{8.6cm}
\includegraphics[width=\textwidth, keepaspectratio]{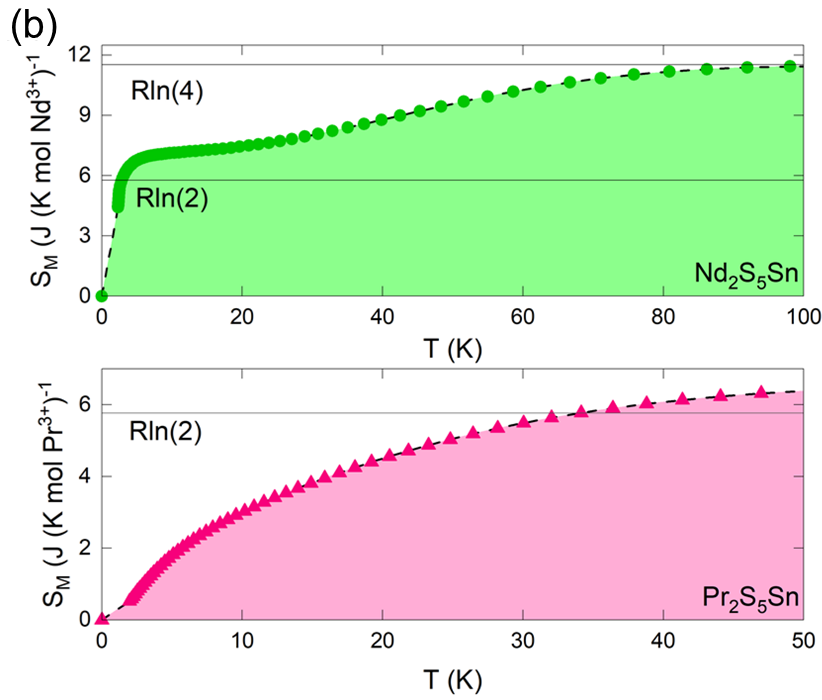}
\caption{\label{fig:fig6b}}
\end{subfigure}{}
\caption{\label{fig:fig6} (a) Heat capacity over temperature (C/T) of La$_2$S$_5$Sn, Nd$_2$S$_5$Sn, and Pr$_2$S$_5$Sn. The inset shows the magnetic heat capacity of Nd$_2$S$_5$Sn and Pr$_2$S$_5$Sn, with the estimated phonon heat capacity subtracted. (b) Magnetic entropy of Nd$_2$S$_5$Sn (top, green) and Pr$_2$S$_5$Sn (bottom, pink), computed by integration of C$_M$/T.}
\end{figure*}

At lower temperatures, three distinct peaks are present in the derivative: one near 0.25\,T, one near 2.2\,T, and one broad peak near 4\,T. These peaks decrease in intensity and shift to lower field as temperature is raised. By 2\,K, the 0.25 and 2.2\,T peaks are not discernable, and the derivative curve appears to have one broad hump centered near 3\,T. This suggests that the loss of anti-ferromagnetic order with increasing field occurs in three steps, with two intermediate states between full AFM order and full alignment with the applied field. The energy difference between the steps decreases with higher temperature.

\subsection{\label{sec:level2}Heat Capacity}

Heat capacity measurements corroborate the magnetization data (Figure~\ref{fig:fig6a}). For Pr$_2$S$_5$Sn, there is a weak divergence of C/T as T\,$\rightarrow$\,0, with no evidence of a phase transition. For Nd$_2$S$_5$Sn, a peak is observed at T\,=\,2.4\,K. Poor fitting of temperature curves below $\approx\,$6\,K by the semi-adiabatic pulse method suggested a first-order phase transition, so a long-pulse technique was used for the low temperature heat capacity. The long-pulse measurements were of larger magnitude near the peak at T\,=\,2\,K, but were in good agreement with the short-pulse data above the peak temperature, consistent with the phase transition being first-order. 

The phonon heat capacity, estimated from the non-magnetic analogue La$_2$S$_5$Sn, was subtracted to find the magnetic contribution (C$_m$). The magnetic entropy was calculated by integrating C$_m$/T (Figure~\ref{fig:fig6b}). For the Nd compound, entropy passes $\Delta$S\,=\,Rln2 near 5\,K, which is sensible given its doublet ground state. It briefly plateaus, and then rises to $\Delta$S\,=\,Rln4 by 100\,K. This is qualitatively consistent but somewhat less than expected from the point charge model, suggesting that the second excited doublet state is somewhat higher in energy than predicted. For Pr, the entropy reaches $\Delta$S\,=\,Rln2 around 35\,K before plateauing, suggesting that only the two lowest-lying energy levels are accessible. The gradual further increase in entropy up to 50\,K is qualitatively consistent with expectations and suggests that the energy gap to the third singlet state is again larger than predicted by the crystal field splitting model. Above $\approx\,$100\,K for the Nd compound and $\approx\,$50\,K for the Pr, the small magnitude of the magnetic heat capacity compared to the subtracted phonon contribution makes the computed entropy unreliable.

\begin{figure}
\includegraphics[width=8.6cm, keepaspectratio]{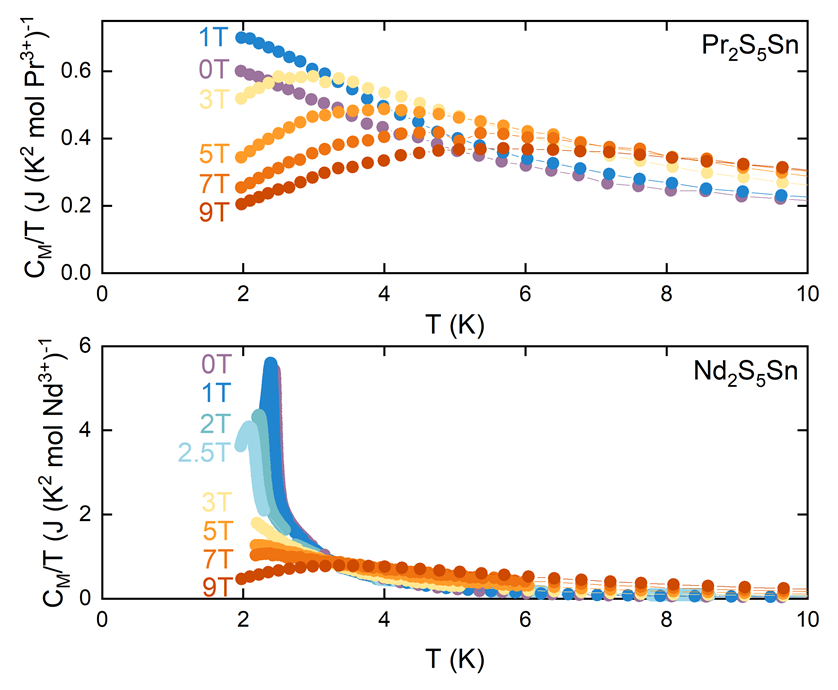}
\caption{\label{fig:fig7}Magnetic heat capacity (as C$_M$/T) of Pr$_2$S$_5$Sn and Nd$_2$S$_5$Sn under applied magnetic fields.}
\end{figure}

\begin{figure}
\includegraphics[width=8.6cm, keepaspectratio]{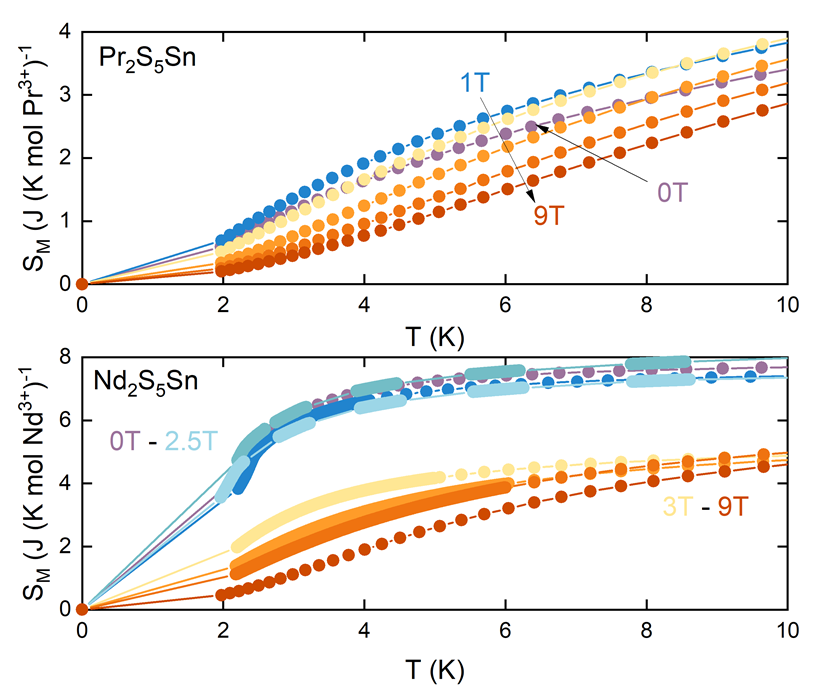}
\caption{\label{fig:fig8}Magnetic entropy of Pr$_2$S$_5$Sn and Nd$_2$S$_5$Sn under applied magnetic field, computed by integration of C$_M$/T.}
\end{figure}

The heat capacity of both compounds was also measured under magnetic field (Figure~\ref{fig:fig7}). For Nd$_2$S$_5$Sn, the 2.5\,K peak is suppresed with field as expected as the presence of a large magnetic field disrupts antiferromagnetic ordering. The peak gradually decreases in magnitude from 0 to 2.5\,T, and seems to disappear completely between 2.5 and 3\,T. This change is clearly visible in the magnetic entropy of these field measurements (Figure~\ref{fig:fig8}). At and above 3\,T, the entropy plateau drops to approximately$\Delta$S\,=\,Rln2, suggesting that some magnetic states are no longer frozen out, or that higher energy states have become inaccessible.

\section{\label{sec:level1}Discussion}

Both our magnetic and thermodynamic measurements make it clear that although they are isostructural, the properties of Pr$_2$S$_5$Sn and Nd$_2$S$_5$Sn are quite distinct. The integer-spin Pr compound is paramagnetic down to at least T\,=\,0.41\,K, while the half-integer Nd compound undergoes an anti-ferromagnetic ordering transition near T\,=\,2.5 K. Besides this most obvious change, we observe that although the two compounds have nearly the same Weiss temperature ($\theta_w$) in Curie-Weiss fits below 30\,K, over the higher temperature range their $\theta_w$ values differ significantly. The crystal field splittings for a point charge model of Nd$^{3+}$ and Pr$^{3+}$ help explain why. The gap between the lowest-lying states (the doublet in Nd and the ``pseudo-doublet'' in Pr) is large compared to the temperature at 30\,K. While both ions effectively have a single doublet state primarily populated, the interaction strengths of the spins in this state may be similar. At higher temperatures where other energy states are accessible, the differences between the two compounds allow the antiferromagnetic exchange in the Nd to become stronger than that of the Pr.

\begin{figure}[b]
\includegraphics[width=8.6cm, keepaspectratio]{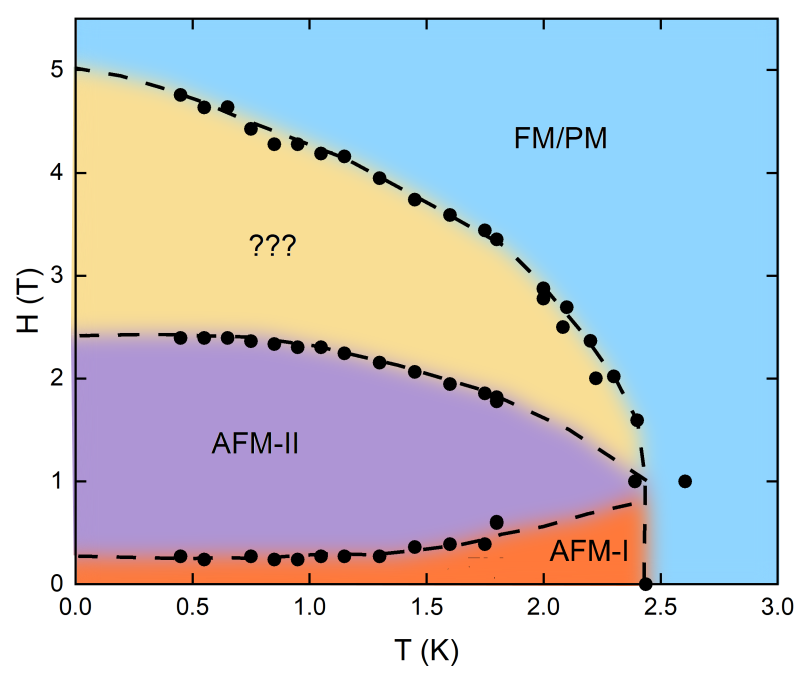}
\caption{\label{fig:fig9}Magnetic phase diagram of Nd$_2$S$_5$Sn, estimated from magnetization and heat capacity measurements}
\end{figure}

A magnetic phase diagram for Nd$_2$S$_5$Sn can be constructed from the M versus H and heat capacity under field results (Figure~\ref{fig:fig9}). From the MvH, we observed that the loss of the antiferromagnetic phase with field occurs in three stages, with the field distance between these stages shrinking at higher temperatures. The two higher-field transitions merge together by about 2\,K, shown in the meeting of the phase boundaries in the diagram. The remaining transition occurs at much lower field ($\approx\,$0.25\,T near 1\,K), indicating a less energetically difficult change in the magnetic order. Since the four phases cannot meet at a single point, one or both of the two lower phase boundaries must cross H\,=\,0\,T below the plotted critical point at T\,=\,2.4\,K, but the temperature where this occurs has not been determined. In the zero-field heat capacity data, the peak appears to be slightly split, suggesting the possibility of two closely-spaced transitions at 2.4\,K. However, as the magnitude of the splitting is only $\approx\,$0.5\,\si{J/(mol.K}), this result is inconclusive. 

How is the magnetic order changing at these metamagnetic transitions? Presumably above the highest field transition, the spins are fully aligned with the applied field, behaving as they would in a ferromagnet. Below this, the specifics of the magnetic order are unknown. The lowest field change may be a spin-flop transition, with spins reorienting to lie parallel to the applied field. In this case, we have one antiferromagnetic ground state and a second intermediate state before all spins align with the field, which is not entirely uncommon in anisotropic antiferromagnets\,\cite{PADUANFILHO, kosaku, MAARTENSE, Tuthill}. The frustration parameter (f\,=\,$\lvert\frac{\theta_w}{T_N}\rvert$) of Nd$_2$S$_5$Sn is f\,$\approx\,$7, relatively low, so the ground state may be a N\'{e}el antiferromagnet. Then at the second transition with field, a subset of the spins may flip along the easy axis, resulting in a magnetic order such as a stripy or zigzag arrangement. Finally, at high enough fields the remaining spins flip to give a ferromagnetic arrangement. However, further study is required to understand the magnetic order. The easy axis is unknown, as the material has only been measured in powder form, and the large deviation of the lattice from equilaterial hexagons may lead to more complicated anisotropic effects.

For Pr$_2$S$_5$Sn, the lack of observed magnetic ordering raises the question of whether it has spin liquid character. Power-law fitting of the field-dependent C$_m$/T versus T did not match the scaling relationship observed in other frustrated spin $\frac{1}{2}$ materials \cite{kimchi}. Additionally, when performing the Curie-Weiss fits, a linear fit was best achieved with the Pauli paramagnetic susceptibility $\chi_0$ equal to zero, in contrast to other candidate QSL materials \cite{balents}. At the same time, the behavior is similar to that observed in other QSL candidates based on Pr, such as Pr$_2$Pb$_2$O$_7$. Further work should clarify the behavior of this candidate.

\section{\label{sec:level1}Conclusion}

We have investigated the physical properties of the isostructural approximate honeycomb compounds Pr$_2$S$_5$Sn and Nd$_2$S$_5$Sn, finding that the Pr compound displays no magnetic ordering down to 0.41\,K, and that the Nd undergoes antiferromagnetic ordering near 2.5\,K. These materials may be usefully compared to the lead pyrochlores Pr$_2$Pb$_2$O$_7$ and Nd$_2$Pb$_2$O$_7$. In these, the Pr material shows no order to 0.4\,K, but has a spin ice like specific heat anomaly at 1.2\,K, which the Nd analog seems to adopt long-range magnetic order at 0.41\,K \cite{hallas}. The similar material Pr$_2$Zr$_2$O$_7$ does not order above 0.2\,K and has excitations consistent with a quantum spin system; like Pr$_2$S$_5$Sn, it has a non-Kramers doublet ground state \cite{Pr2Zr2O7, Kimura2013}. Pr$_2$S$_5$Sn lacks order at low temperatures and the frustration parameter f $\approx\,$18\,$>$\,10 suggests significant magnetic frustration. Understanding how this frustration occurs on the geometry of this approximate honeycomb is of interest. Nd$_2$S$_5$Sn displays a series of magnetic transitions under applied field, and seems to adopt an intermediate magnetic order between its AFM and FM states. Neutron scattering measurements on this compound to determine the magnetic order, and lower-temperature characterization of Pr$_2$S$_5$Sn, would allow us to better understand these materials.

\begin{acknowledgments}
This work was supported as part of the Institute for Quantum Matter, and Energy Frontier Research Center funded by the U.S. Department of Energy, Office of Science, Office of Basic Energy Sciences, under Award DE-SC0019331. The $^3$He MPMS was funded by the National Science Foundation, Division of Materials Research, Major Research Instrumentation Program, under Award 1828490.
\end{acknowledgments}

%\appendix

%add table with refinement parameters

%\section{}

%\nocite{*}

\bibliographystyle{apsrev4-2} 
\bibliography{Nd2S5Sn}% Produces the bibliography via BibTeX.

\newcommand{\noopsort}[1]{} \newcommand{\printfirst}[2]{#1}
  \newcommand{\singleletter}[1]{#1} \newcommand{\switchargs}[2]{#2#1}
\begin{thebibliography}{35}%
\makeatletter
\providecommand \@ifxundefined [1]{%
 \@ifx{#1\undefined}
}%
\providecommand \@ifnum [1]{%
 \ifnum #1\expandafter \@firstoftwo
 \else \expandafter \@secondoftwo
 \fi
}%
\providecommand \@ifx [1]{%
 \ifx #1\expandafter \@firstoftwo
 \else \expandafter \@secondoftwo
 \fi
}%
\providecommand \natexlab [1]{#1}%
\providecommand \enquote  [1]{``#1''}%
\providecommand \bibnamefont  [1]{#1}%
\providecommand \bibfnamefont [1]{#1}%
\providecommand \citenamefont [1]{#1}%
\providecommand \href@noop [0]{\@secondoftwo}%
\providecommand \href [0]{\begingroup \@sanitize@url \@href}%
\providecommand \@href[1]{\@@startlink{#1}\@@href}%
\providecommand \@@href[1]{\endgroup#1\@@endlink}%
\providecommand \@sanitize@url [0]{\catcode `\\12\catcode `\$12\catcode
  `\&12\catcode `\#12\catcode `\^12\catcode `\_12\catcode `\%12\relax}%
\providecommand \@@startlink[1]{}%
\providecommand \@@endlink[0]{}%
\providecommand \url  [0]{\begingroup\@sanitize@url \@url }%
\providecommand \@url [1]{\endgroup\@href {#1}{\urlprefix }}%
\providecommand \urlprefix  [0]{URL }%
\providecommand \Eprint [0]{\href }%
\providecommand \doibase [0]{https://doi.org/}%
\providecommand \selectlanguage [0]{\@gobble}%
\providecommand \bibinfo  [0]{\@secondoftwo}%
\providecommand \bibfield  [0]{\@secondoftwo}%
\providecommand \translation [1]{[#1]}%
\providecommand \BibitemOpen [0]{}%
\providecommand \bibitemStop [0]{}%
\providecommand \bibitemNoStop [0]{.\EOS\space}%
\providecommand \EOS [0]{\spacefactor3000\relax}%
\providecommand \BibitemShut  [1]{\csname bibitem#1\endcsname}%
\let\auto@bib@innerbib\@empty
%</preamble>
\bibitem [{\citenamefont {Balents}(2010)}]{balents}%
  \BibitemOpen
  \bibfield  {author} {\bibinfo {author} {\bibfnamefont {L.}~\bibnamefont
  {Balents}},\ }\href@noop {} {\bibfield  {journal} {\bibinfo  {journal}
  {Nature}\ }\textbf {\bibinfo {volume} {464}},\ \bibinfo {pages} {199}
  (\bibinfo {year} {2010})}\BibitemShut {NoStop}%
\bibitem [{\citenamefont {Broholm}\ \emph {et~al.}(2020)\citenamefont
  {Broholm}, \citenamefont {Cava}, \citenamefont {Kivelson}, \citenamefont
  {Nocera}, \citenamefont {Norman},\ and\ \citenamefont {Senthil}}]{broholm}%
  \BibitemOpen
  \bibfield  {author} {\bibinfo {author} {\bibfnamefont {C.}~\bibnamefont
  {Broholm}}, \bibinfo {author} {\bibfnamefont {R.~J.}\ \bibnamefont {Cava}},
  \bibinfo {author} {\bibfnamefont {S.~A.}\ \bibnamefont {Kivelson}}, \bibinfo
  {author} {\bibfnamefont {D.~G.}\ \bibnamefont {Nocera}}, \bibinfo {author}
  {\bibfnamefont {M.~R.}\ \bibnamefont {Norman}},\ and\ \bibinfo {author}
  {\bibfnamefont {T.}~\bibnamefont {Senthil}},\ }\href
  {https://science.sciencemag.org/content/367/6475/eaay0668} {\bibfield
  {journal} {\bibinfo  {journal} {Science}\ }\textbf {\bibinfo {volume} {367}}
  (\bibinfo {year} {2020})}\BibitemShut {NoStop}%
\bibitem [{\citenamefont {Chamorro}\ \emph {et~al.}(2020)\citenamefont
  {Chamorro}, \citenamefont {McQueen},\ and\ \citenamefont {Tran}}]{jcqsl}%
  \BibitemOpen
  \bibfield  {author} {\bibinfo {author} {\bibfnamefont {J.~R.}\ \bibnamefont
  {Chamorro}}, \bibinfo {author} {\bibfnamefont {T.~M.}\ \bibnamefont
  {McQueen}},\ and\ \bibinfo {author} {\bibfnamefont {T.~T.}\ \bibnamefont
  {Tran}},\ }\href@noop {} {\bibfield  {journal} {\bibinfo  {journal} {Chem.
  Rev.}\ } (\bibinfo {year} {2020})}\BibitemShut {NoStop}%
\bibitem [{\citenamefont {Kitaev}(2006)}]{kitaev}%
  \BibitemOpen
  \bibfield  {author} {\bibinfo {author} {\bibfnamefont {A.}~\bibnamefont
  {Kitaev}},\ }\href@noop {} {\bibfield  {journal} {\bibinfo  {journal} {Ann.
  Phys.}\ }\textbf {\bibinfo {volume} {321}} (\bibinfo {year}
  {2006})}\BibitemShut {NoStop}%
\bibitem [{\citenamefont {{Sears}}\ \emph {et~al.}(2015)\citenamefont
  {{Sears}}, \citenamefont {{Songvilay}}, \citenamefont {{Plumb}},
  \citenamefont {{Clancy}}, \citenamefont {{Qiu}}, \citenamefont {{Zhao}},
  \citenamefont {{Parshall}},\ and\ \citenamefont {{Kim}}}]{sears}%
  \BibitemOpen
  \bibfield  {author} {\bibinfo {author} {\bibfnamefont {J.~A.}\ \bibnamefont
  {{Sears}}}, \bibinfo {author} {\bibfnamefont {M.}~\bibnamefont
  {{Songvilay}}}, \bibinfo {author} {\bibfnamefont {K.~W.}\ \bibnamefont
  {{Plumb}}}, \bibinfo {author} {\bibfnamefont {J.~P.}\ \bibnamefont
  {{Clancy}}}, \bibinfo {author} {\bibfnamefont {Y.}~\bibnamefont {{Qiu}}},
  \bibinfo {author} {\bibfnamefont {Y.}~\bibnamefont {{Zhao}}}, \bibinfo
  {author} {\bibfnamefont {D.}~\bibnamefont {{Parshall}}},\ and\ \bibinfo
  {author} {\bibfnamefont {Y.-J.}\ \bibnamefont {{Kim}}},\ }\href
  {https://doi.org/10.1103/PhysRevB.91.144420} {\bibfield  {journal} {\bibinfo
  {journal} {\prb}\ }\textbf {\bibinfo {volume} {91}},\ \bibinfo {eid} {144420}
  (\bibinfo {year} {2015})}\BibitemShut {NoStop}%
\bibitem [{\citenamefont {Chaloupka}\ \emph {et~al.}(2010)\citenamefont
  {Chaloupka}, \citenamefont {Jackeli},\ and\ \citenamefont
  {Khaliullin}}]{chal}%
  \BibitemOpen
  \bibfield  {author} {\bibinfo {author} {\bibfnamefont {J.}~\bibnamefont
  {Chaloupka}}, \bibinfo {author} {\bibfnamefont {G.}~\bibnamefont {Jackeli}},\
  and\ \bibinfo {author} {\bibfnamefont {G.}~\bibnamefont {Khaliullin}},\
  }\href {https://doi.org/10.1103/PhysRevLett.105.027204} {\bibfield  {journal}
  {\bibinfo  {journal} {Phys. Rev. Lett.}\ }\textbf {\bibinfo {volume} {105}},\
  \bibinfo {pages} {027204} (\bibinfo {year} {2010})}\BibitemShut {NoStop}%
\bibitem [{\citenamefont {Singh}\ \emph {et~al.}(2012)\citenamefont {Singh},
  \citenamefont {Manni}, \citenamefont {Reuther}, \citenamefont {Berlijn},
  \citenamefont {Thomale}, \citenamefont {Ku}, \citenamefont {Trebst},\ and\
  \citenamefont {Gegenwart}}]{singh}%
  \BibitemOpen
  \bibfield  {author} {\bibinfo {author} {\bibfnamefont {Y.}~\bibnamefont
  {Singh}}, \bibinfo {author} {\bibfnamefont {S.}~\bibnamefont {Manni}},
  \bibinfo {author} {\bibfnamefont {J.}~\bibnamefont {Reuther}}, \bibinfo
  {author} {\bibfnamefont {T.}~\bibnamefont {Berlijn}}, \bibinfo {author}
  {\bibfnamefont {R.}~\bibnamefont {Thomale}}, \bibinfo {author} {\bibfnamefont
  {W.}~\bibnamefont {Ku}}, \bibinfo {author} {\bibfnamefont {S.}~\bibnamefont
  {Trebst}},\ and\ \bibinfo {author} {\bibfnamefont {P.}~\bibnamefont
  {Gegenwart}},\ }\href {https://doi.org/10.1103/PhysRevLett.108.127203}
  {\bibfield  {journal} {\bibinfo  {journal} {Phys. Rev. Lett.}\ }\textbf
  {\bibinfo {volume} {108}},\ \bibinfo {pages} {127203} (\bibinfo {year}
  {2012})}\BibitemShut {NoStop}%
\bibitem [{\citenamefont {Stavropoulos}\ \emph {et~al.}(2019)\citenamefont
  {Stavropoulos}, \citenamefont {Pereira},\ and\ \citenamefont {Kee}}]{stavro}%
  \BibitemOpen
  \bibfield  {author} {\bibinfo {author} {\bibfnamefont {P.~P.}\ \bibnamefont
  {Stavropoulos}}, \bibinfo {author} {\bibfnamefont {D.}~\bibnamefont
  {Pereira}},\ and\ \bibinfo {author} {\bibfnamefont {H.-Y.}\ \bibnamefont
  {Kee}},\ }\href {https://doi.org/10.1103/PhysRevLett.123.037203} {\bibfield
  {journal} {\bibinfo  {journal} {Phys. Rev. Lett.}\ }\textbf {\bibinfo
  {volume} {123}},\ \bibinfo {pages} {037203} (\bibinfo {year}
  {2019})}\BibitemShut {NoStop}%
\bibitem [{\citenamefont {Dong}\ and\ \citenamefont {Sheng}(2020)}]{dong}%
  \BibitemOpen
  \bibfield  {author} {\bibinfo {author} {\bibfnamefont {X.-Y.}\ \bibnamefont
  {Dong}}\ and\ \bibinfo {author} {\bibfnamefont {D.~N.}\ \bibnamefont
  {Sheng}},\ }\href {https://doi.org/10.1103/PhysRevB.102.121102} {\bibfield
  {journal} {\bibinfo  {journal} {Phys. Rev. B}\ }\textbf {\bibinfo {volume}
  {102}},\ \bibinfo {pages} {121102(R)} (\bibinfo {year} {2020})}\BibitemShut
  {NoStop}%
\bibitem [{\citenamefont {Li}\ \emph {et~al.}(2016)\citenamefont {Li},
  \citenamefont {Bishop},\ and\ \citenamefont {Campbell}}]{bishop}%
  \BibitemOpen
  \bibfield  {author} {\bibinfo {author} {\bibfnamefont {P.~H.~Y.}\
  \bibnamefont {Li}}, \bibinfo {author} {\bibfnamefont {R.~F.}\ \bibnamefont
  {Bishop}},\ and\ \bibinfo {author} {\bibfnamefont {C.~E.}\ \bibnamefont
  {Campbell}},\ }\href@noop {} {\bibfield  {journal} {\bibinfo  {journal} {J.
  Phys.: Conf. Ser.}\ }\textbf {\bibinfo {volume} {702}},\ \bibinfo {pages}
  {021001} (\bibinfo {year} {2016})}\BibitemShut {NoStop}%
\bibitem [{\citenamefont {Rau}\ \emph {et~al.}(2014)\citenamefont {Rau},
  \citenamefont {Lee},\ and\ \citenamefont {Kee}}]{rau}%
  \BibitemOpen
  \bibfield  {author} {\bibinfo {author} {\bibfnamefont {J.~G.}\ \bibnamefont
  {Rau}}, \bibinfo {author} {\bibfnamefont {E.~K.-H.}\ \bibnamefont {Lee}},\
  and\ \bibinfo {author} {\bibfnamefont {H.-Y.}\ \bibnamefont {Kee}},\
  }\href@noop {} {\bibfield  {journal} {\bibinfo  {journal} {Phys. Rev. Lett.}\
  }\textbf {\bibinfo {volume} {112}},\ \bibinfo {pages} {077204} (\bibinfo
  {year} {2014})}\BibitemShut {NoStop}%
\bibitem [{\citenamefont {Rao}\ \emph {et~al.}(2014)\citenamefont {Rao},
  \citenamefont {Sankar}, \citenamefont {Muthuselvam},\ and\ \citenamefont
  {Chou}}]{rao}%
  \BibitemOpen
  \bibfield  {author} {\bibinfo {author} {\bibfnamefont {G.~N.}\ \bibnamefont
  {Rao}}, \bibinfo {author} {\bibfnamefont {R.}~\bibnamefont {Sankar}},
  \bibinfo {author} {\bibfnamefont {I.~P.}\ \bibnamefont {Muthuselvam}},\ and\
  \bibinfo {author} {\bibfnamefont {F.~C.}\ \bibnamefont {Chou}},\ }\href@noop
  {} {\bibfield  {journal} {\bibinfo  {journal} {J. Magn. Magn. Mater.}\
  }\textbf {\bibinfo {volume} {370}},\ \bibinfo {pages} {13} (\bibinfo {year}
  {2014})}\BibitemShut {NoStop}%
\bibitem [{\citenamefont {Venderbos}\ \emph {et~al.}(2011)\citenamefont
  {Venderbos}, \citenamefont {Daghofer}, \citenamefont {van~den Brink},\ and\
  \citenamefont {Kumar}}]{venderbos}%
  \BibitemOpen
  \bibfield  {author} {\bibinfo {author} {\bibfnamefont {J.~W.~F.}\
  \bibnamefont {Venderbos}}, \bibinfo {author} {\bibfnamefont {M.}~\bibnamefont
  {Daghofer}}, \bibinfo {author} {\bibfnamefont {J.}~\bibnamefont {van~den
  Brink}},\ and\ \bibinfo {author} {\bibfnamefont {S.}~\bibnamefont {Kumar}},\
  }\href@noop {} {\bibfield  {journal} {\bibinfo  {journal} {Phys. Rev. Lett.}\
  }\textbf {\bibinfo {volume} {107}},\ \bibinfo {pages} {076405} (\bibinfo
  {year} {2011})}\BibitemShut {NoStop}%
\bibitem [{\citenamefont {Bordelon}\ \emph {et~al.}(2020)\citenamefont
  {Bordelon}, \citenamefont {Liu}, \citenamefont {Posthuma}, \citenamefont
  {Sarte}, \citenamefont {Butch}, \citenamefont {Pajerowski}, \citenamefont
  {Banerjee}, \citenamefont {Balents},\ and\ \citenamefont {Wilson}}]{NaYbO2}%
  \BibitemOpen
  \bibfield  {author} {\bibinfo {author} {\bibfnamefont {M.~M.}\ \bibnamefont
  {Bordelon}}, \bibinfo {author} {\bibfnamefont {C.}~\bibnamefont {Liu}},
  \bibinfo {author} {\bibfnamefont {L.}~\bibnamefont {Posthuma}}, \bibinfo
  {author} {\bibfnamefont {P.~M.}\ \bibnamefont {Sarte}}, \bibinfo {author}
  {\bibfnamefont {N.~P.}\ \bibnamefont {Butch}}, \bibinfo {author}
  {\bibfnamefont {D.~M.}\ \bibnamefont {Pajerowski}}, \bibinfo {author}
  {\bibfnamefont {A.}~\bibnamefont {Banerjee}}, \bibinfo {author}
  {\bibfnamefont {L.}~\bibnamefont {Balents}},\ and\ \bibinfo {author}
  {\bibfnamefont {S.~D.}\ \bibnamefont {Wilson}},\ }\href
  {https://doi.org/10.1103/PhysRevB.101.224427} {\bibfield  {journal} {\bibinfo
   {journal} {Phys. Rev. B}\ }\textbf {\bibinfo {volume} {101}},\ \bibinfo
  {pages} {224427} (\bibinfo {year} {2020})}\BibitemShut {NoStop}%
\bibitem [{\citenamefont {Li}\ \emph {et~al.}(2015)\citenamefont {Li},
  \citenamefont {Chen}, \citenamefont {Tong}, \citenamefont {Pi}, \citenamefont
  {Liu}, \citenamefont {Yang}, \citenamefont {Wang},\ and\ \citenamefont
  {Zhang}}]{YbMgGaO4}%
  \BibitemOpen
  \bibfield  {author} {\bibinfo {author} {\bibfnamefont {Y.}~\bibnamefont
  {Li}}, \bibinfo {author} {\bibfnamefont {G.}~\bibnamefont {Chen}}, \bibinfo
  {author} {\bibfnamefont {W.}~\bibnamefont {Tong}}, \bibinfo {author}
  {\bibfnamefont {L.}~\bibnamefont {Pi}}, \bibinfo {author} {\bibfnamefont
  {J.}~\bibnamefont {Liu}}, \bibinfo {author} {\bibfnamefont {Z.}~\bibnamefont
  {Yang}}, \bibinfo {author} {\bibfnamefont {X.}~\bibnamefont {Wang}},\ and\
  \bibinfo {author} {\bibfnamefont {Q.}~\bibnamefont {Zhang}},\ }\href
  {https://doi.org/10.1103/PhysRevLett.115.167203} {\bibfield  {journal}
  {\bibinfo  {journal} {Phys. Rev. Lett.}\ }\textbf {\bibinfo {volume} {115}},\
  \bibinfo {pages} {167203} (\bibinfo {year} {2015})}\BibitemShut {NoStop}%
\bibitem [{\citenamefont {Gaudet}\ \emph {et~al.}(2019)\citenamefont {Gaudet},
  \citenamefont {Smith}, \citenamefont {Dudemaine}, \citenamefont {Beare},
  \citenamefont {Buhariwalla}, \citenamefont {Butch}, \citenamefont {Stone},
  \citenamefont {Kolesnikov}, \citenamefont {Xu}, \citenamefont {Yahne},
  \citenamefont {Ross}, \citenamefont {Marjerrison}, \citenamefont {Garrett},
  \citenamefont {Luke}, \citenamefont {Bianchi},\ and\ \citenamefont
  {Gaulin}}]{Ce2Zr2O7}%
  \BibitemOpen
  \bibfield  {author} {\bibinfo {author} {\bibfnamefont {J.}~\bibnamefont
  {Gaudet}}, \bibinfo {author} {\bibfnamefont {E.~M.}\ \bibnamefont {Smith}},
  \bibinfo {author} {\bibfnamefont {J.}~\bibnamefont {Dudemaine}}, \bibinfo
  {author} {\bibfnamefont {J.}~\bibnamefont {Beare}}, \bibinfo {author}
  {\bibfnamefont {C.~R.~C.}\ \bibnamefont {Buhariwalla}}, \bibinfo {author}
  {\bibfnamefont {N.~P.}\ \bibnamefont {Butch}}, \bibinfo {author}
  {\bibfnamefont {M.~B.}\ \bibnamefont {Stone}}, \bibinfo {author}
  {\bibfnamefont {A.~I.}\ \bibnamefont {Kolesnikov}}, \bibinfo {author}
  {\bibfnamefont {G.}~\bibnamefont {Xu}}, \bibinfo {author} {\bibfnamefont
  {D.~R.}\ \bibnamefont {Yahne}}, \bibinfo {author} {\bibfnamefont {K.~A.}\
  \bibnamefont {Ross}}, \bibinfo {author} {\bibfnamefont {C.~A.}\ \bibnamefont
  {Marjerrison}}, \bibinfo {author} {\bibfnamefont {J.~D.}\ \bibnamefont
  {Garrett}}, \bibinfo {author} {\bibfnamefont {G.~M.}\ \bibnamefont {Luke}},
  \bibinfo {author} {\bibfnamefont {A.~D.}\ \bibnamefont {Bianchi}},\ and\
  \bibinfo {author} {\bibfnamefont {B.~D.}\ \bibnamefont {Gaulin}},\ }\href
  {https://doi.org/10.1103/PhysRevLett.122.187201} {\bibfield  {journal}
  {\bibinfo  {journal} {Phys. Rev. Lett.}\ }\textbf {\bibinfo {volume} {122}},\
  \bibinfo {pages} {187201} (\bibinfo {year} {2019})}\BibitemShut {NoStop}%
\bibitem [{\citenamefont {Wen}\ \emph {et~al.}(2017)\citenamefont {Wen},
  \citenamefont {Koohpayeh}, \citenamefont {Ross}, \citenamefont {Trump},
  \citenamefont {McQueen}, \citenamefont {Kimura}, \citenamefont {Nakatsuji},
  \citenamefont {Qiu}, \citenamefont {Pajerowski}, \citenamefont {Copley},\
  and\ \citenamefont {Broholm}}]{Pr2Zr2O7}%
  \BibitemOpen
  \bibfield  {author} {\bibinfo {author} {\bibfnamefont {J.-J.}\ \bibnamefont
  {Wen}}, \bibinfo {author} {\bibfnamefont {S.~M.}\ \bibnamefont {Koohpayeh}},
  \bibinfo {author} {\bibfnamefont {K.~A.}\ \bibnamefont {Ross}}, \bibinfo
  {author} {\bibfnamefont {B.~A.}\ \bibnamefont {Trump}}, \bibinfo {author}
  {\bibfnamefont {T.~M.}\ \bibnamefont {McQueen}}, \bibinfo {author}
  {\bibfnamefont {K.}~\bibnamefont {Kimura}}, \bibinfo {author} {\bibfnamefont
  {S.}~\bibnamefont {Nakatsuji}}, \bibinfo {author} {\bibfnamefont
  {Y.}~\bibnamefont {Qiu}}, \bibinfo {author} {\bibfnamefont {D.~M.}\
  \bibnamefont {Pajerowski}}, \bibinfo {author} {\bibfnamefont {J.~R.~D.}\
  \bibnamefont {Copley}},\ and\ \bibinfo {author} {\bibfnamefont {C.~L.}\
  \bibnamefont {Broholm}},\ }\href
  {https://doi.org/10.1103/PhysRevLett.118.107206} {\bibfield  {journal}
  {\bibinfo  {journal} {Phys. Rev. Lett.}\ }\textbf {\bibinfo {volume} {118}},\
  \bibinfo {pages} {107206} (\bibinfo {year} {2017})}\BibitemShut {NoStop}%
\bibitem [{\citenamefont {Toby}\ and\ \citenamefont {von Dreele}(2013)}]{gsas}%
  \BibitemOpen
  \bibfield  {author} {\bibinfo {author} {\bibfnamefont {B.}~\bibnamefont
  {Toby}}\ and\ \bibinfo {author} {\bibfnamefont {R.}~\bibnamefont {von
  Dreele}},\ }\href@noop {} {\bibfield  {journal} {\bibinfo  {journal} {J.
  Appl. Crystallogr.}\ }\textbf {\bibinfo {volume} {46}},\ \bibinfo {pages}
  {544} (\bibinfo {year} {2013})}\BibitemShut {NoStop}%
\bibitem [{\citenamefont {Momma}\ and\ \citenamefont {Izumi}(2011)}]{vesta}%
  \BibitemOpen
  \bibfield  {author} {\bibinfo {author} {\bibfnamefont {K.}~\bibnamefont
  {Momma}}\ and\ \bibinfo {author} {\bibfnamefont {F.}~\bibnamefont {Izumi}},\
  }\href@noop {} {\bibfield  {journal} {\bibinfo  {journal} {J. Appl.
  Crystallogr.}\ }\textbf {\bibinfo {volume} {44}},\ \bibinfo {pages} {1272}
  (\bibinfo {year} {2011})}\BibitemShut {NoStop}%
\bibitem [{\citenamefont {Scheie}(2020)}]{pycrystalfield}%
  \BibitemOpen
  \bibfield  {author} {\bibinfo {author} {\bibfnamefont {A.}~\bibnamefont
  {Scheie}},\ }\href@noop {} {\bibfield  {journal} {\bibinfo  {journal} {arXiv
  preprint}\ }\textbf {\bibinfo {volume} {2006.15151}} (\bibinfo {year}
  {2020})}\BibitemShut {NoStop}%
\bibitem [{\citenamefont {Scheie}(2018)}]{LPHC}%
  \BibitemOpen
  \bibfield  {author} {\bibinfo {author} {\bibfnamefont {A.}~\bibnamefont
  {Scheie}},\ }\href@noop {} {\bibfield  {journal} {\bibinfo  {journal} {J. Low
  Temp. Phys.}\ }\textbf {\bibinfo {volume} {193}},\ \bibinfo {pages} {60}
  (\bibinfo {year} {2018})}\BibitemShut {NoStop}%
\bibitem [{\citenamefont {Jaulmes}(1974)}]{jaulmes}%
  \BibitemOpen
  \bibfield  {author} {\bibinfo {author} {\bibfnamefont {P.}~\bibnamefont
  {Jaulmes}},\ }\href@noop {} {\bibfield  {journal} {\bibinfo  {journal} {Acta
  Crystallogr. Sect. B}\ ,\ \bibinfo {pages} {73+}} (\bibinfo {year}
  {1974})}\BibitemShut {NoStop}%
\bibitem [{\citenamefont {Daszkiewicz}\ \emph {et~al.}(2008)\citenamefont
  {Daszkiewicz}, \citenamefont {Gulay},\ and\ \citenamefont {Shemet}}]{dasz}%
  \BibitemOpen
  \bibfield  {author} {\bibinfo {author} {\bibfnamefont {M.}~\bibnamefont
  {Daszkiewicz}}, \bibinfo {author} {\bibfnamefont {L.}~\bibnamefont {Gulay}},\
  and\ \bibinfo {author} {\bibfnamefont {V.}~\bibnamefont {Shemet}},\
  }\href@noop {} {\bibfield  {journal} {\bibinfo  {journal} {Acta Crystallogr.
  Sect. B}\ ,\ \bibinfo {pages} {172}} (\bibinfo {year} {2008})}\BibitemShut
  {NoStop}%
\bibitem [{\citenamefont {Bain}\ and\ \citenamefont {Berry}(2008)}]{bain}%
  \BibitemOpen
  \bibfield  {author} {\bibinfo {author} {\bibfnamefont {G.~A.}\ \bibnamefont
  {Bain}}\ and\ \bibinfo {author} {\bibfnamefont {J.~F.}\ \bibnamefont
  {Berry}},\ }\href@noop {} {\bibfield  {journal} {\bibinfo  {journal} {J.
  Chem. Ed.}\ }\textbf {\bibinfo {volume} {85}},\ \bibinfo {pages} {532}
  (\bibinfo {year} {2008})}\BibitemShut {NoStop}%
\bibitem [{\citenamefont {Hallas}\ \emph {et~al.}(2015)\citenamefont {Hallas},
  \citenamefont {Arevalo-Lopez}, \citenamefont {Sharma}, \citenamefont
  {Munsie}, \citenamefont {Attfield}, \citenamefont {Wiebe},\ and\
  \citenamefont {Luke}}]{hallas}%
  \BibitemOpen
  \bibfield  {author} {\bibinfo {author} {\bibfnamefont {A.~M.}\ \bibnamefont
  {Hallas}}, \bibinfo {author} {\bibfnamefont {A.~M.}\ \bibnamefont
  {Arevalo-Lopez}}, \bibinfo {author} {\bibfnamefont {A.~Z.}\ \bibnamefont
  {Sharma}}, \bibinfo {author} {\bibfnamefont {T.}~\bibnamefont {Munsie}},
  \bibinfo {author} {\bibfnamefont {J.~P.}\ \bibnamefont {Attfield}}, \bibinfo
  {author} {\bibfnamefont {C.~R.}\ \bibnamefont {Wiebe}},\ and\ \bibinfo
  {author} {\bibfnamefont {G.~M.}\ \bibnamefont {Luke}},\ }\href
  {https://doi.org/10.1103/PhysRevB.91.104417} {\bibfield  {journal} {\bibinfo
  {journal} {Phys. Rev. B}\ }\textbf {\bibinfo {volume} {91}},\ \bibinfo
  {pages} {104417} (\bibinfo {year} {2015})}\BibitemShut {NoStop}%
\bibitem [{\citenamefont {Kimura}\ \emph {et~al.}(2013)\citenamefont {Kimura},
  \citenamefont {Nakatsuji}, \citenamefont {Wen}, \citenamefont {Broholm},
  \citenamefont {Stone}, \citenamefont {Nishibori},\ and\ \citenamefont
  {Sawa}}]{Kimura2013}%
  \BibitemOpen
  \bibfield  {author} {\bibinfo {author} {\bibfnamefont {K.}~\bibnamefont
  {Kimura}}, \bibinfo {author} {\bibfnamefont {S.}~\bibnamefont {Nakatsuji}},
  \bibinfo {author} {\bibfnamefont {J.-J.}\ \bibnamefont {Wen}}, \bibinfo
  {author} {\bibfnamefont {C.}~\bibnamefont {Broholm}}, \bibinfo {author}
  {\bibfnamefont {M.~B.}\ \bibnamefont {Stone}}, \bibinfo {author}
  {\bibfnamefont {E.}~\bibnamefont {Nishibori}},\ and\ \bibinfo {author}
  {\bibfnamefont {H.}~\bibnamefont {Sawa}},\ }\href
  {https://doi.org/10.1038/ncomms2914} {\bibfield  {journal} {\bibinfo
  {journal} {Nat. Commun.s}\ }\textbf {\bibinfo {volume} {4}},\ \bibinfo
  {pages} {1934} (\bibinfo {year} {2013})}\BibitemShut {NoStop}%
\bibitem [{\citenamefont {Princep}\ \emph {et~al.}(2013)\citenamefont
  {Princep}, \citenamefont {Prabhakaran}, \citenamefont {Boothroyd},\ and\
  \citenamefont {Adroja}}]{princep}%
  \BibitemOpen
  \bibfield  {author} {\bibinfo {author} {\bibfnamefont {A.~J.}\ \bibnamefont
  {Princep}}, \bibinfo {author} {\bibfnamefont {D.}~\bibnamefont
  {Prabhakaran}}, \bibinfo {author} {\bibfnamefont {A.~T.}\ \bibnamefont
  {Boothroyd}},\ and\ \bibinfo {author} {\bibfnamefont {D.~T.}\ \bibnamefont
  {Adroja}},\ }\href {https://doi.org/10.1103/PhysRevB.88.104421} {\bibfield
  {journal} {\bibinfo  {journal} {Phys. Rev. B}\ }\textbf {\bibinfo {volume}
  {88}},\ \bibinfo {pages} {104421} (\bibinfo {year} {2013})}\BibitemShut
  {NoStop}%
\bibitem [{\citenamefont {Matsuhira}\ \emph {et~al.}(2002)\citenamefont
  {Matsuhira}, \citenamefont {Hinatsu}, \citenamefont {Tenya}, \citenamefont
  {Amitsuka},\ and\ \citenamefont {Sakakibara}}]{matsuhira}%
  \BibitemOpen
  \bibfield  {author} {\bibinfo {author} {\bibfnamefont {K.}~\bibnamefont
  {Matsuhira}}, \bibinfo {author} {\bibfnamefont {Y.}~\bibnamefont {Hinatsu}},
  \bibinfo {author} {\bibfnamefont {K.}~\bibnamefont {Tenya}}, \bibinfo
  {author} {\bibfnamefont {H.}~\bibnamefont {Amitsuka}},\ and\ \bibinfo
  {author} {\bibfnamefont {T.}~\bibnamefont {Sakakibara}},\ }\href
  {https://doi.org/10.1143/JPSJ.71.1576} {\bibfield  {journal} {\bibinfo
  {journal} {J. Phys. Soc. Japan}\ }\textbf {\bibinfo {volume} {71}},\ \bibinfo
  {pages} {1576} (\bibinfo {year} {2002})}\BibitemShut {NoStop}%
\bibitem [{\citenamefont {Ciomaga~Hatnean}\ \emph {et~al.}(2014)\citenamefont
  {Ciomaga~Hatnean}, \citenamefont {Decorse}, \citenamefont {Lees},
  \citenamefont {Petrenko}, \citenamefont {Keeble},\ and\ \citenamefont
  {Balakrishnan}}]{CiomagaHatnean2014}%
  \BibitemOpen
  \bibfield  {author} {\bibinfo {author} {\bibfnamefont {M.}~\bibnamefont
  {Ciomaga~Hatnean}}, \bibinfo {author} {\bibfnamefont {C.}~\bibnamefont
  {Decorse}}, \bibinfo {author} {\bibfnamefont {M.~R.}\ \bibnamefont {Lees}},
  \bibinfo {author} {\bibfnamefont {O.~A.}\ \bibnamefont {Petrenko}}, \bibinfo
  {author} {\bibfnamefont {D.~S.}\ \bibnamefont {Keeble}},\ and\ \bibinfo
  {author} {\bibfnamefont {G.}~\bibnamefont {Balakrishnan}},\ }\href
  {https://doi.org/10.1088/2053-1591/1/2/026109} {\bibfield  {journal}
  {\bibinfo  {journal} {Mater. Res. Express}\ }\textbf {\bibinfo {volume}
  {1}},\ \bibinfo {pages} {026109} (\bibinfo {year} {2014})}\BibitemShut
  {NoStop}%
\bibitem [{\citenamefont {Bertin}\ \emph {et~al.}(2015)\citenamefont {Bertin},
  \citenamefont {Dalmas~de R\'eotier}, \citenamefont {F\aa{}k}, \citenamefont
  {Marin}, \citenamefont {Yaouanc}, \citenamefont {Forget}, \citenamefont
  {Sheptyakov}, \citenamefont {Frick}, \citenamefont {Ritter}, \citenamefont
  {Amato}, \citenamefont {Baines},\ and\ \citenamefont {King}}]{Bertin}%
  \BibitemOpen
  \bibfield  {author} {\bibinfo {author} {\bibfnamefont {A.}~\bibnamefont
  {Bertin}}, \bibinfo {author} {\bibfnamefont {P.}~\bibnamefont {Dalmas~de
  R\'eotier}}, \bibinfo {author} {\bibfnamefont {B.}~\bibnamefont {F\aa{}k}},
  \bibinfo {author} {\bibfnamefont {C.}~\bibnamefont {Marin}}, \bibinfo
  {author} {\bibfnamefont {A.}~\bibnamefont {Yaouanc}}, \bibinfo {author}
  {\bibfnamefont {A.}~\bibnamefont {Forget}}, \bibinfo {author} {\bibfnamefont
  {D.}~\bibnamefont {Sheptyakov}}, \bibinfo {author} {\bibfnamefont
  {B.}~\bibnamefont {Frick}}, \bibinfo {author} {\bibfnamefont
  {C.}~\bibnamefont {Ritter}}, \bibinfo {author} {\bibfnamefont
  {A.}~\bibnamefont {Amato}}, \bibinfo {author} {\bibfnamefont
  {C.}~\bibnamefont {Baines}},\ and\ \bibinfo {author} {\bibfnamefont
  {P.~J.~C.}\ \bibnamefont {King}},\ }\href@noop {} {\bibfield  {journal}
  {\bibinfo  {journal} {{Phys. Rev. B}}\ }\textbf {\bibinfo {volume} {{92}}},\
  \bibinfo {pages} {144423} (\bibinfo {year} {{2015}})}\BibitemShut {NoStop}%
\bibitem [{\citenamefont {Paduan-Filho}\ \emph {et~al.}(1986)\citenamefont
  {Paduan-Filho}, \citenamefont {Droiche},\ and\ \citenamefont
  {Becerra}}]{PADUANFILHO}%
  \BibitemOpen
  \bibfield  {author} {\bibinfo {author} {\bibfnamefont {A.}~\bibnamefont
  {Paduan-Filho}}, \bibinfo {author} {\bibfnamefont {M.~S.}\ \bibnamefont
  {Droiche}},\ and\ \bibinfo {author} {\bibfnamefont {C.~C.}\ \bibnamefont
  {Becerra}},\ }\href
  {https://doi.org/https://doi.org/10.1016/0304-8853(86)90216-7} {\bibfield
  {journal} {\bibinfo  {journal} {J. Magn. Magn. Mater.}\ }\textbf {\bibinfo
  {volume} {54-57}},\ \bibinfo {pages} {699} (\bibinfo {year}
  {1986})}\BibitemShut {NoStop}%
\bibitem [{\citenamefont {Yamada}\ and\ \citenamefont
  {Kanamori}(1967)}]{kosaku}%
  \BibitemOpen
  \bibfield  {author} {\bibinfo {author} {\bibfnamefont {K.}~\bibnamefont
  {Yamada}}\ and\ \bibinfo {author} {\bibfnamefont {J.}~\bibnamefont
  {Kanamori}},\ }\href {https://doi.org/10.1143/PTP.38.541} {\bibfield
  {journal} {\bibinfo  {journal} {Prog. Theo. Phys.}\ }\textbf {\bibinfo
  {volume} {38}},\ \bibinfo {pages} {541} (\bibinfo {year} {1967})}\BibitemShut
  {NoStop}%
\bibitem [{\citenamefont {Maartense}\ \emph {et~al.}(1977)\citenamefont
  {Maartense}, \citenamefont {Yaeger},\ and\ \citenamefont
  {Wanklyn}}]{MAARTENSE}%
  \BibitemOpen
  \bibfield  {author} {\bibinfo {author} {\bibfnamefont {I.}~\bibnamefont
  {Maartense}}, \bibinfo {author} {\bibfnamefont {I.}~\bibnamefont {Yaeger}},\
  and\ \bibinfo {author} {\bibfnamefont {B.~M.}\ \bibnamefont {Wanklyn}},\
  }\href {https://doi.org/https://doi.org/10.1016/0038-1098(77)91485-5}
  {\bibfield  {journal} {\bibinfo  {journal} {Solid State Commun.}\ }\textbf
  {\bibinfo {volume} {21}},\ \bibinfo {pages} {93} (\bibinfo {year}
  {1977})}\BibitemShut {NoStop}%
\bibitem [{\citenamefont {Tuthill}(1981)}]{Tuthill}%
  \BibitemOpen
  \bibfield  {author} {\bibinfo {author} {\bibfnamefont {G.~F.}\ \bibnamefont
  {Tuthill}},\ }\href {https://doi.org/10.1088/0022-3719/14/18/016} {\bibfield
  {journal} {\bibinfo  {journal} {J. Phys. C}\ }\textbf {\bibinfo {volume}
  {14}},\ \bibinfo {pages} {2483} (\bibinfo {year} {1981})}\BibitemShut
  {NoStop}%
\bibitem [{\citenamefont {Kimchi}\ \emph {et~al.}(2018)\citenamefont {Kimchi},
  \citenamefont {Sheckelton}, \citenamefont {McQueen},\ and\ \citenamefont
  {Lee}}]{kimchi}%
  \BibitemOpen
  \bibfield  {author} {\bibinfo {author} {\bibfnamefont {I.}~\bibnamefont
  {Kimchi}}, \bibinfo {author} {\bibfnamefont {J.~P.}\ \bibnamefont
  {Sheckelton}}, \bibinfo {author} {\bibfnamefont {T.~M.}\ \bibnamefont
  {McQueen}},\ and\ \bibinfo {author} {\bibfnamefont {P.}~\bibnamefont {Lee}},\
  }\href@noop {} {\bibfield  {journal} {\bibinfo  {journal} {Nat. Commun.}\
  }\textbf {\bibinfo {volume} {22}},\ \bibinfo {pages} {9(1):4367} (\bibinfo
  {year} {2018})}\BibitemShut {NoStop}%
\end{thebibliography}%

\end{document}